# A Hybrid Approach using Ontology Similarity and Fuzzy Logic for Semantic Question Answering

Monika Rani∗, Maybin K. Muyeba, and O. P. Vyas

Indian Institute of Information Technology, Allahabad, INDIA.
School of Computing, Mathematics and Digital Technology Manchester Metropolitan Univ., U.K.
Indian Institute of Information Technology, Allahabad, INDIA.

**Abstract.** One of the challenges in information retrieval is providing accurate answers to a user's question often expressed as uncertainty words. Most answers are based on a Syntactic approach rather than a Semantic analysis of the query. In this paper our objective is to present a hybrid approach for a Semantic question answering retrieval system using Ontology Similarity and Fuzzy logic. We use a Fuzzy co-clustering algorithm to retrieve collection of documents based on Ontology Similarity. Fuzzy scale uses Fuzzy type-1 for documents and Fuzzy type-2 for words to prioritize answers. The objective of this work is to provide retrieval systems with more accurate answers than non-fuzzy Semantic Ontology approach.

## 1 Introduction

The Educational Semantic web aims to discover knowledge using educational learning areas such as personal learning, education administration and knowledge construction [12] Semantic web (web 3.0) is providing data integrity capabilities by not only machine readability but also machine analysis. Education is improving by using Semantic web approaches where a large number of online students share data semantically. In addition, student portals help students to be connected everywhere. Electronic textbooks [1] provide open context from sources like openstax, ck12.org, crowd sourc-ing, NCERT etc. Massive open online courses (Moocs) for example coursera, udacity, khan, Edx, TED-Ed also small virtual classes are easily found on Internet.. The web is naturally fuzzy in nature, so text document and building Ontology requires a fuzzy approach. To improve the education Semantic web, the first step is a semantic question answering system where uncertain words are questions. To implement such as a system, a Fuzzy Ontology approach can be applied by using fuzzy logic (type-1 or type-2 levels) [11] for text retrieval. A fuzzy scale is proposed for two levels, first for membership of document (Fuzzy type-1) and membership of the word, where the words have uncertainty (by way of synonyms, as Fuzzy type-2). A fuzzy co-clustering algorithm is used to simultaneously cluster documents and words and hence handle the overlapping nature of documents in terms of membership functions. To get even more meaningful results, Semantic Fuzzy Ontology is proposed as a hybrid approach for question answering. The question answering system is based on semantic approaches as well as ontology driven representation of knowledge. The rest of the paper is organized as follows: section 2 gives a background; section 3 presents a methodology including comparisons and use of Fuzzy type-2, and a conclusion in section 4.

## 2 Background

### 2.1 Question and Answering System

In information retrieval, the challenge is to find accurate answers to questions asked by the user. Questionnaire Mining helps to give accurate answers by handling complex words for which Fuzzy type-2 and linguistic variables can be considered. Thus a Fuzzy Ontology Information Retrieval System (FOIRS) [4] can play a vital role in understanding semantic relationships. FOIRS provides the basis to find co-relationships between user query terms with the document terms. The user query analysis can be done in much the same way as syntax analysis and also as semantic analysis for question answering. If the user wants to search any information which is already present in the database for example, if our digital library stores information about painting created by Ravi on subjects such as Irises, nature, soil etc. and the user queries the database with conjunctions between the keywords "Painting" and "Ravi" and "Irises", then no accurate results will be returned as keywords are not enough basis to reach an accurate answer. A semantic system considers the structure of sentences as a set of objects, functions and various relationships between them. For the user input query example "Painting" by "Ravi" with a subject type "Blossom", a semantic query system will retrieve an accurate result even the query terms vary, for example "Painting" by "Ravi" with subject "Irises". This is because the ontology defines "Blossom" as subclass-hierarchy of Irises in the Knowledge graph representation. The ontology plays a vital role in understanding such ambiguous user questions and helps retrieve appropriate answers. Ontology is way toward semantic analysis for the question answering search engines like Google, Yahoo etc. For these purposes, ontology indexing ensemble with semantic relations among terms is useful.

The question answering (QA) systems main challenge is to retrieve accurate answers to questions [7] asked by users not only based on keywords, but also on semantic bases, summarized by various approaches:

- Syntax query based retrieval
- FAQ (question templates) based retrieval
- Semantic query based retrieval
- Ontology based retrieval
- Transparent query based

WordNet and link grammar approaches toward scaling QA [5] for the web can prove helpful tools in recommender systems and feedback analysis.

## 2.2 Ontology

Ontology plays an important role in development of knowledge based systems to describe semantic relationships among entities. Ontologies basically describe a formal conceptualization of a domain of interest. Fuzzy Ontology [6] can help in understanding semantic relationships by applying fuzzy logic to deal with the vagueness of data. Fuzzy type-1 can deal with crisp membership, whereas Fuzzy type-2 deals with fuzziness of fuzzy membership (See Table 1). Scientifically, using Fuzzy type-1 set as a model for words is incorrect as it is unable to deal with uncertainty because words mean different things to different people in nature [10]. This uncertainty about words can be further classified into two types:

- Intra uncertainty: This is uncertainty that a person has about the word.
- Inter uncertainty: This is uncertainty that a group of people have about the word.

Table 1: Comparison of Fuzzy type-1 and Fuzzy type-2.

| Fuzzy type-1 | Fuzzy type-2 |
| --- | --- |
| Level 1 | Level 2 |
| Membership Document | Membership of words (synonyms) |
| Uncertainty is in range [0,1] | Uncertainty is measured by an additional dimension |
| Two dimension | Three dimension |
| Notation use A | Notation use tilde A |

The proposed methodology uses fuzzy concepts like linguistic variable and Fuzzy type-2 for information retrieval. Fuzzy type-2 models can deal with the uncertainty of words. Fuzzy type-2 reduces to Fuzzy type-1 in cases where there is no uncertainty in the scenario. Ontologies play important role in Information extraction.

Ontologies represent knowledge in a graph conceptual diagram using semantic approach rather than syntactic approach where each node shows either a document or a word. Various ontologies match a user query and finally retrieves the ontology for the query as knowledge based (short-path), corpus based (co-occurrence), information content and probability of encountering an instance. Then ontology matching is used as a solution to the semantic heterogeneity problem. Applying reasoning from an ontology to text data play an important role in question answering systems.

For Ontology Similarity, an edge count method can be used for calculating similarity [2] between a keyword question and hierarchical ontology tree to obtain semantic relations. For two similar words, the return value is 1 and 0 for two dissimilar words, defined by the equation:

$$St(t1, t2) = (e^{xd} - 1)/(e^{xd} + e^{ys} - 2)$$

where d = depth of tree, S= shorted path length, x and y are smoothing factors and $(t1, t2)$ is the similarity value ranging from 0 to 1.

Protégé OWL plug-in [3] shows a major change in describing information of various ontologies by adding new facilities. OWL ontology can be categorized as OWL Lite, OWL Full and OWL DL [3]. OWL DL can be considered as the extension of OWL Lite. Similarly OWL Full is an extension to OWL DL. Semantic web uses RDB2onto, DB2OWL and check d2rq etc to match between ontology and database. Ontologies do not only represent lexical knowledge, but complex world knowledge about events. Ontologies can be created by Protégé tools, software and after that, use Protégé Java API or translate the ontology into a rule base using Fuxi [8].

### 2.3 Data Clustering

In hard clustering, data elements are partitioned in such a way that any single data element can belong to only one cluster rather than to many clusters. Fuzzy clustering [9] represents data elements that partition data in such a way that data can belong to two or more clusters with the degree of belongingness overlapping between the cluster.

## 3 Methodology Description

The user enters a source string as a question. The first objective of the machine is to syntactically analyze the text from the source. Only after that the Lexical Analysis can be done for each term in a question and then tokenized by removing stop words present in the user's question. The next step is linguistic preprocessing; POS (part of speech) are tagged in such a way that Syntactic analysis can be done easily as shown in Fig. 1. In POS, a tree is created to differentiate between each question term and label. Each term is labeled as a noun, a verb or adjective. The Structural sequence is identified by POS. Then questions can be interpreted for its semantic meaning. WordNet tool shows the results for all available synonyms for nouns and verbs. This tool represents knowledge which is also useful for creating a lexical ontology for the domain knowledge. A word can be processed semantically by WordNet tools. The groups of words describing the same intension are called synsets. The edge-count method is used to match for question similarity with the existing ontology. Fuzzy co-cluster is used to present collection of answers and fuzzy scale (Fuzzy type-1 for document and Fuzzy type-2 for words) in order to score the collection obtained by fuzzy co-clustering. The final result is the matrix where the x-axis represents "Ontology Similarity" and the y-axis represents "keywords".

Our proposed algorithm is as follows:

a) Input text in search engine (Question).
b) Parse the question for structural analysis.
c) Remove stop words for keyword extraction.
d) Use WordNet to get synonyms of a word in the keyword. Generate all possible combinations of synonyms
e) Retrieval is based on the semantic ontology similarity (edge-count method) match for questions; where the question is matched with the answer on the basis of existing ontologies.
f) The result is obtained from the matrix where the x-axis represents "Ontology Similarity" and y-axis represents "keywords".
g) Use fuzzy co-cluster to retrieve answers by using semantic ontology similarity.
h) Retrieve the final answer from the matrix by prioritizing answers obtained by fuzzy co-clustering using a fuzzy scale.

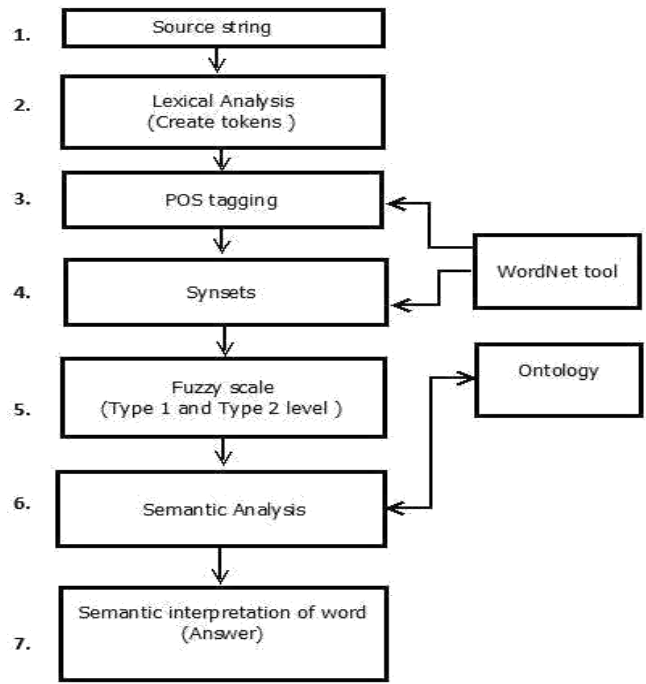

**Fig. 1.** Flow diagram of Semantic question answer

Fuzzy co-clustering manages data and features into two or more clusters at the same point of time. Note that the overlapping structure of web documents is represented in the cluster with the degree of belongingness for each web document. Reasons to choose the proposed fuzzy co-clustering in our case are:

a) Fuzzy co-clustering is a technique to manage cluster data (Document) and features (Words) [9] into two or more clusters at the same point of time. Here, co-clustering (or bi-clustering) has the ability to capture overlap between web documents and words mentioned in the documents. The degree of belongingness for each document and word are mentioned in co-clustering.

b) The fuzzy co-clustering has the following advantages over the traditional clustering:
   • Dimensionality reduction as the feature is stored in overlapping form for various clusters.
   • Fuzzy co-clustering provides efficient results in situations which are vague and uncertain.
   • Interpretability of document clusters becomes easy.
   • Improvement in accuracy due to local model of clustering.
   • Fuzzy membership functions improve representation of overlapping clusters in answers by using semantic ontology similarity.

c) Fuzzy type-2 deals with 3-D (three dimensional data) while FCC_STF [9] algorithm has the ability to deal with the curse of dimensionality and outliers.

d) Fuzzy co-clustering concept is used in algorithms like FCCM, Fuzzy codok and FCC-STF as describe in Table 2. FCC_STF is found to be the best in comparison to FCCM and Fuzzy codok with the new single term fuzzifier approach. FCC_STF is a solution to the curse of dimensionality and outliers.

Table 2: Comparison of Co-clustering Algorithm.

| Categories | FCCM | Fuzzy Codok | FCC_STF |
|---|---|---|---|
| Algorithm for co-clustering | Fuzzy co-clustering for categorical multi-variate | Fuzzy co-clustering of document and keywords | Fuzzy co-clustering with single term fuzzifier |
| Fuzzifier | Fuzzy entropy is use as Fuzzifier in FCCM Algorithm | Fuzzy gini index is used as fuzzifier in fuzzy codok Algorithm | Single term fuzzifier is used in FCC_STF algorithm |
| Advantage | Algorithm for co-lustering | Ability to deal with the exponential problem | Clipping for negative value and renormalization take place |
| Disadvantage | Overflow (exponential) problem | Negative membership | |

To retrieve accurate answers, semantic processing plays an important role. Fuzzy scale (Fig. 2) is an approach towards the semantic analysis of the question at levels 1 and 2. Level 1 represents the membership of document in a cluster ($\mu = 0.7$). Here Fuzzy type-1 is used for the document as it unable to deal with uncertainty, whereas Level 2 represents the membership of word in a cluster ($\mu = 0.61 - 0.69$) using Fuzzy type-2. As one word can have different meanings to different users, uncertainty comes into play. Fuzzy type-2 [10] has ability to deal with uncertainty which can be helpful to deal with the synonyms present in the user question, while Fuzzy type-1 considers no uncertainty.

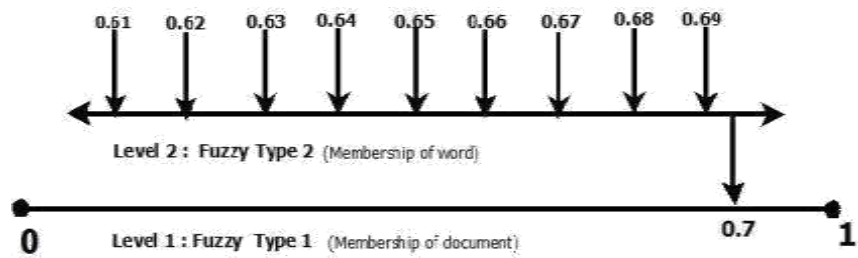

**Fig. 2.** Fuzzy Scale

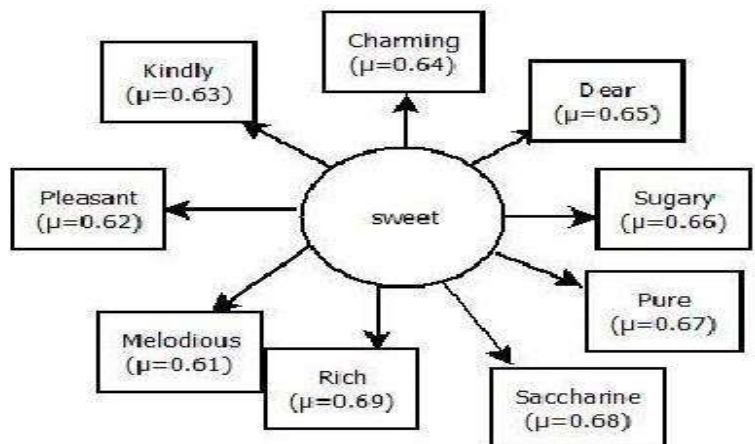

**Fig. 3.** Membership of word (Fuzzy type)

Calculating the score:

Score = (Membership of document (A) + Membership of Word (˜A))/Number of documents = ($\mu$(doc A)+ $\mu$(word(˜A)/N). From Fig. 3, the upper membership function for the word is $\mu = 0,69$, and the lower membership function is $\mu = 0.61$.

Fuzzy type-2 is used for computation of the word as it has the ability to deal with linguistic uncertainty. Whereas Fuzzy type-1 has crisp membership like for document ($\mu = 0.7$), Fuzzy type-2 has a fuzzy membership for synonymous words ($\mu = 0.61 - 0.69$), it can be called fuzzy-fuzzy set. Here the computation of word is applied to find appropriate synonym for each question. An exact synonym helps in obtaining the meaning of the question. So to retrieve appropriate answers, semantic analysis of each query term along with synonyms is a must.

From Fig. 3, "sweet" is a vague term which we use every day in common language. The term sweet depends on perception based assessment. The same word "sweet" has different meanings. When a user types the term "sweet" in the search engine as a question, this term is treated as a vague term. But uncertainty arises in associating the word "sweet" particularly to sugar. Here uncertainty can arises because the term "sweet" can be associated to describe behaviors like kind, melodious, musical not only to the sugar. In Fig. 3, various memberships of word "sweet" are described. Let us consider the following statements where the term "sweet" needs to be checked for a similar context with respect to its meaning, for which Fuzzy linguistic rules can be applied.

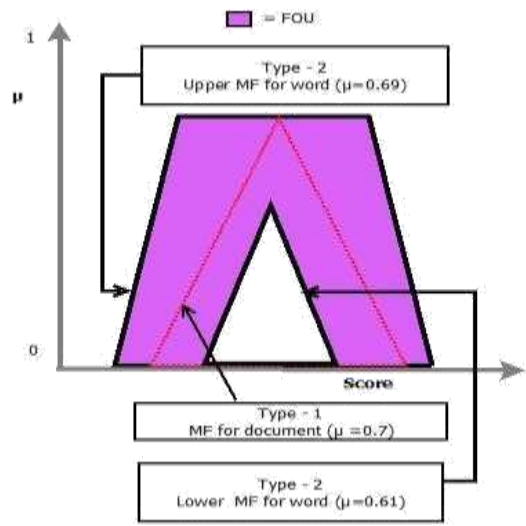

**Fig. 4.** Foundation of uncertainty (FOU)

For example:

a) "Sarah" is such a sweet little girl. She's always looking after her brother" -
   Kindly ($\mu = 0.63$).
b) "This tea is too sweet for me to drink, how much sugar is in it?" – Sugary ($\mu = 0.66$).

Fuzzy type-2 can be visualized by plotting footprints of uncertainty (FOU) in a 2-D domain representation form as shown in Fig. 4. Fuzzy type-2 represents three dimensions of data whereas Fuzzy type-1 represents two dimensional data. The uniform color represents the uniformity of possibilities. Due to this uniformity, Fuzzy type-2 is called Interval type-2 represented by IT2. Till now, there is not much progress in IT2 as it's unable to choose best secondary membership functions, but computation of words has been an emerging field well placed to use this aspect.

## 4 Conclusion and Future Work

We have proposed a hybrid approach for Semantic question answering based on Semantic Fuzzy ontology for retrieval systems. Fuzzy co-clustering is used to retrieve the answers by matching user's question with the existing hierarchical ontology. A fuzzy scale is used to prioritize the answers retrieved by matrix using Fuzzy co-clustering. For Fuzzy co-clustering, the FCC_STF algorithm is preferred to FCCM and Fuzzy codok. Future work will implement this hybrid approach based on the proposed semantic fuzzy ontology with various applications including e-learning and intelligent web search systems. Users not only get syntactic answers, but also semantic answers based on the question terms. The proposed question answering system provides a gateway for deep web search along with surface web search.